\documentclass[pra,showkeys,twocolumn,showpacs]{revtex4}
\usepackage[english]{babel}
\usepackage{amsmath}
\usepackage{amsfonts}
\usepackage{graphicx}

\newcommand{\abs}[1]{\left|#1\right|}

\newcommand{\cc}[1]{{{#1}^\ast}}

\begin{document}

\title{A simple renormalization group approximation of the groundstate properties of interacting bosonic systems}

\date{19th March, 2007}

\author{Roman Werpachowski}

\email[Corresponding author: ]{roman@cft.edu.pl}

\affiliation{Center for Theoretical Physics PAS, Al. Lotnik\'{o}w 32/46, 02-668 Warsaw, Poland}

\author{Jerzy Kijowski}

\affiliation{Center for Theoretical Physics PAS, Al. Lotnik\'{o}w 32/46, 02-668 Warsaw, Poland}

\affiliation{Cardinal Stefan Wyszy\'{n}ski University in Warsaw, Poland}

\begin{abstract}
We present a new, simple renormalization group method of investigating groundstate properties of interacting bosonic systems. Our method reduces the number of particles in a system, which makes numerical calculations possible for large systems. It is conceptually simple and easy to implement, and allows to investigate the properties unavailable through mean field approximations, such as one- and two-particle reduced density matrices of the groundstate. As an example, we model a weakly interacting 1D Bose gas in a harmonic trap. Compared to the mean-field Gross-Pitaevskii approximation, our method provides a more accurate description of the groundstate one-particle density matrix. We have also obtained the Hall-Post lower bounds for the groundstate energy of the gas. All results have been obtained by the straightforward numerical diagonalization of the Hamiltonian matrix.
\end{abstract}



\maketitle

\section{Introduction}
\label{sec:introduction}

The numerical investigation of the groundstate properties of a multiparticle interacting bosonic system is a much harder task than in the case of a single particle system. The naive approach consists in choosing a large enough finite Hilbert space basis and the numerical diagonalization of the resulting Hamiltonian matrix. However, the necessary basis size grows exponentially with the number of particles, which makes this simple method inadequate for the treatment of large systems. To avoid this problem, many approximations have been invented, such as the Gross-Pitaevskii (GP) mean field approach~\cite{pitaevskii2,pitstr}, the Density Matrix Renormalization Group (DMRG) method~\cite{white1,schollwock} or, in the case of strong interactions, the Thomas-Fermi approximation~\cite{pitstr} and the Tonks-Girardeau model~\cite{tonks,girardeau}. For fermionic systems, the Exact Diagonalization-Ab Initio (EDABI) method has been implemented~\cite{spalek1,spalek2,spalek3}. One should also note the exceptional case of a full analytical solution in one dimension by Lieb and Liniger~\cite{lieb}. From this solution, two- and three-pair correlation functions of an interacting one-dimensional (1D) Bose gas have been derived~\cite{shlyapnikov,shlyapnikov2,shlyapnikov3}. In this paper we present a new approach, which is in the flavour of Renormalization Group methods but is conceptually simple and easy to implement. Our method amends the problem of unmanageable basis size by reducing the number of particles in the system and renormalizing the Hamiltonian. We approximate one- and two-particle properties of the large system with the same properties of the smaller system. In contrast to mean field methods, our approach allows to calculate such quantities as one- or two-particle reduced density matrices (1-RDM and 2-RDM) of the groundstate.

In Section~\ref{sec:method}, we describe how our method works. Section~\ref{sec:example} contains an example application of the method to the problem of one-dimensional interacting Bose gas in a harmonic trap. The results are summarized in Section~\ref{sec:summary}.

\section{Hamiltonian renormalization and the approximation of groundstate properties}
\label{sec:method}

We present our approximation in the case of a system with two-body interactions. It can be easily generalized to the general case of $n$-body interactions.

Consider a 1D Hamiltonian describing a system of $N$ scalar (zero spin) bosons,
\begin{equation}
\hat{H}_N = \sum_{k=1}^N \left( - \frac{\hbar^2}{2m} \frac{\partial^2}{\partial x_k^2} + V_1(x_k) \right) + \sum_{k=1}^N \sum_{k'=1}^{k-1} V_2(x_k, x_{k'}) \ ,
\end{equation}
where $N$ is the number of particles, $m$ is the particle mass, $V_1(x)$ is the external one-particle potential and $V_2(x,x') = V_2(x',x)$ is the two-particle interaction potential.

Investigating such a system, we are often interested only in one- or two-particle properties of the groundstate. One way to calculate them is to obtain an approximation of the 1-RDM or 2-RDM of the groundstate.

We approximate the system by replacing it with a much smaller system containing $N' \ll N$ scalar bosons. The smaller system is governed by a renormalized version of the original Hamiltonian $\hat{H}_N$, that is
\begin{equation}
\label{eq:HNprim}
\begin{split}
\hat{H}_{N'} &= \frac{N}{N'} \Biggl[ \sum_{k=1}^{N'} \left( - \frac{\hbar^2}{2m} \frac{\partial^2}{\partial x_k^2} + V_1(x_k) \right)\\
&\quad+ \frac{N-1}{N'-1} \sum_{k=1}^{N'} \sum_{k'=1}^{k-1} V_2(x_k, x_{k'}) \Biggr] \ .
\end{split}
\end{equation}
We calculate the properties of interest from the groundstate of $\hat{H}_{N'}$, thus avoiding the insurmountable problem of diagonalizing the Hamiltonian of the large system, $\hat{H}_N$. The results for increasing values of $N'$ will converge to the values of the corresponding properties of the $N$-particle system.

We will now justify our procedure. Let $\Psi$ and $\rho$ be a state of the large system  and its $N'$-RDM, respectively. It is easy to show that their mean energies, measured by the respective Hamiltonians, are equal,
\begin{equation}
\label{eq:aven}
\mathrm{Tr} \left( \hat{H}_{N'} \rho \right) = \left\langle \Psi \left| \hat{H}_N \right| \Psi \right\rangle \ .
\end{equation}
%Hence, if the mean energy of an $N$-particle state is lower than the mean energy of some other $N$-particle state, then the mean energies of their $N'$-RDMs are related in the same way. 
Hence, when the mean energy of $\Psi$ becomes lower, moving closer to the mean energy of the groundstate $\Psi_0$ of $\hat{H}_N$, the mean energy of $\rho$, also becomes lower and moves closer to the mean energy of the (pure state) density matrix $\rho'_0$ of the groundstate $\Psi'_0$ of $\hat{H}_{N'}$. Because of the variational principle, the density matrix $\rho'_0$ is an approximation of the reduced density matrix (RDM) $\rho_0$ of the groundstate $\Psi_0$. The one- or two-particle properties of $\rho_0$ (i.e. of $\Psi_0$), like the probability density, are approximated by the same properties of $\rho'_0$ (i.e. of $\Psi'_0$). Since $N' \ll N$, it is much easier to calculate numerically the groundstate $\Psi'_0$ than the groundstate $\Psi_0$, and to investigate the one- or two- particle properties of $\Psi_0$ by investigating the same properties of $\Psi'_0$.

The main source of error in our method is the fact that the variational search for the groundstate converges to the $N'$-particle groundstate $\Psi_0'$, not to the RDM of the $N$-particle groundstate, $\rho_0$. This is because for bosons, not every $N'$-particle density matrix is an RDM of an $N$-particle state. A better strategy would be to perform the variational search not in the whole $N'$-particle Hilbert space but in the smaller space of $N'$-particle density matrices which are RDMs of $N$-particle states. However, the problem of identifying this space, the so-called $N$-representability problem~\cite{coleman,ruskai,gidofalvi}, remains unsolved. Therefore, we have to perform our calculations for a sequence of $N'$. The energy of the $N'$-particle groundstate is a lower bound of the energy of the investigated $N$-particle groundstate~\cite{hallpost}. When $N'$ increases, the groundstate energy increases and approaches the groundstate energy of the $N$-particle groundstate. Due to the variational principle, this means that the $N'$-particle groundstates approximate the $N$-particle groundstate increasingly well, in the sense that the one- and two-particle properties calculated from these groundstates converge to the corresponding properties of the $N$-particle groundstate.

In the general case of $n$-particle interactions, the renormalization goes as follows: an $n$-particle interaction potential term $\sum_{k_1=1}^{N} \sum_{k_2=1}^{k_1-1} \dotsm \sum_{k_2=1}^{k_1-1} \dotsm \sum_{k_n = 1}^{k_{n-1}-1} V_n(x_1,\dotsc,x_n)$,
symmetrical with respect to permutations of coordinates $x_k$, is replaced by the term $\frac{N(N-1) \cdot \dotsm \cdot (N-n+1)}{N'(N'-1) \cdot \dotsm \cdot (N'-n+1)}$ $ \sum_{k_1=1}^{N'} \sum_{k_2=1}^{k_1-1} \dotsm \sum_{k_n = 1}^{k_{n-1}-1} V_n(x_1,\dotsc,x_n)$. Equation~\eqref{eq:aven} is true also in this general case.

\section{A simple example}
\label{sec:example}

\subsection{Investigated system}

In our example, we consider a system of $N=100$ scalar bosons with a dimensionless Hamiltonian
\[
\hat{H}_N = \sum_{k=1}^N \left( - \frac{1}{2} \frac{\partial^2}{\partial x_k^2} + \frac{1}{2} x^2_k \right) + \lambda \sum_{k=1}^N \sum_{k'=1}^{k-1} \delta(x_k - x_{k'}) \ ,
\]
where $\delta(x-x')$ is the Dirac $\delta$ function. This interaction potential is often used to describe cold bosons forming a Bose-Einstein condensate (BEC) in a trap, when only the $s$-wave scattering occurs~\cite{pitstr}. Our example concerns positive $\lambda$, which lead to the repulsive interaction.

We have approximated numerically the 1-RDM and 2-RDM of the groundstate for two values of interaction strength $\lambda$. The procedure begins with the calculation of a finite matrix of the renormalized Hamiltonian $\hat{H}_{N'}$ in a finite basis composed of the noninteracting Hamiltonian ($\lambda = 0$) eigenstates, permanents~\footnote{A permanent is the bosonic counterpart of the Slater determinant. For example, for two particles the wavefunction $\Psi(x_1,x_2) \sim \psi_a(x_1) \psi_b(x_2) + \psi_b(x_1) \psi_a(x_2)$ is the permanent of one-particle wavefunctions $\psi_a$ and $\psi_b$.} of $N'$ one-particle Hermite functions $\mathcal{H}_k$,
\begin{equation}
\label{eq:Hermite}
\mathcal{H}_k(x) = \frac{1}{\sqrt{k! 2^k \sqrt{\pi}}} H_k(x) \exp\left( - \frac{x^2}{2} \right) \ ,
\end{equation}
where $H_k(x)$ is $k$-th Hermite polynomial. The basis contains all eigenstates with (nonrenormalized) energies lower than a cutoff energy $L + N'/2$, i.e. permanents of functions $\mathcal{H}_{k_n}$, $n=1,\dotsc,N'$ such that $\sum_{n=1}^{N'} k_n < L$. Then, the groundstate is calculated with the help of an iterative Lanczos-type numerical procedure~\cite{arpack}. From the groundstate we obtain the 1-RDM and the 2-RDM, with trace normalized to unity. Using them, we can calculate any one- or two-particle property of the groundstate. For given $N'$, the basis cutoff $L$ is chosen to be large enough so that calculated properties do not change upon further increase of $L$.

In the case of 1-RDM, $\rho_1$, we compare the diagonal part, $\rho_1(x,x)$, with the probability density calculated by minimizing numerically the GP energy functional~\cite{pitstr} of our system,
\begin{equation}
\label{eq:GPen}
\begin{split}
E[\Psi] &= \frac{N}{2} \int_{-\infty}^\infty \Biggl[ \cc{\Psi}(x) \left( - \frac{\partial^2}{\partial x^2} + x^2 \right) \Psi(x)\\
&\quad+ (N-1) \lambda \abs{\Psi(x)}^4 \Biggr] \mathrm{d} x \ .
\end{split}
\end{equation}
The minimization is performed by expanding the wavefunction $\Psi(x)$ in the finite basis of the first twenty Hermite functions~\eqref{eq:Hermite}, inserting the expansion into~\eqref{eq:GPen} and minimizing numerically the resulting functional of the expansion coefficients.

We present numerical results for two values of $\lambda$, $10^{-2}$ and $5 \cdot 10^{-2}$. Both values are in the regime of weak interactions, where the minimization of the GP energy functional provides a good approximation of the groundstate.

\subsection{Groundstate energy}

First, we provide the data on the groundstate energy $E_0$. In the noninteracting case $\lambda = 0$, $E_0$ is precisely known and equals 50. Table~\ref{tab:en} lists three different approximations of $E_0$ for two nonzero values of $\lambda$. In the second and third column of Table~\ref{tab:en}, two different upper bounds for $E_0$ are listed: the one obtained from the GP functional, $E_\mathrm{GP}$, and the variational bound, $E_\mathrm{Gauss}$, calculated as a minimal mean value of $\hat{H}_N$ in a state $\Psi_\sigma$, a product of $N$ Gaussian one-particle wavefunctions with a common variational parameter $\sigma$,
\[
%\Psi_\sigma(x) = \frac{1}{\sqrt[4]{2\pi\sigma^2}} \exp \left( - \frac{x^2}{4 \sigma^2}  \right)
\Psi_\sigma(x_1,\dotsc,x_N) = \prod_{k=1}^N \frac{1}{\sqrt[4]{2\pi\sigma^2}} \exp \left( - \frac{x^2_k}{4 \sigma^2}  \right) \ ,
\]
i.e. $E_\mathrm{Gauss} = \min_{\sigma \in \mathbb{R}} \langle \Psi_\sigma | \hat{H}_N | \Psi_\sigma \rangle$. Relatively small differences between $E_\mathrm{GP}$ and $E_\mathrm{Gauss}$ indicate that the groundstates are close to Gaussian. As expected, the GP approximation provides a better estimation of the groundstate energy than the Gaussian \textit{Ansatz}. Our method provides an estimation of the true groundstate energy $E_0$ as the so-called Hall-Post~\cite{hallpost} lower bound, $E_\mathrm{HP}$. Values of $E_\mathrm{HP}$ are listed in the fourth column of Table~\ref{tab:en}.
\begin{table}[h!]
\begin{center}
\begin{tabular}{|r|r|r|r|}
\hline
$\lambda$ & $E_\mathrm{GP}$ & $E_\mathrm{Gauss}$ & $E_\mathrm{HP}$\\
\hline
$10^{-2}$ & 68.8 & 68.9 & 68.2 \\
$5 \cdot 10^{-2}$ & 130.9 & 132.3 & 121.4\\
\hline
\end{tabular}
\end{center}
\caption{Bounds for the groundstate energy $E_0$. The first two, $E_\mathrm{GP}$ and $E_\mathrm{Gauss}$, are the upper bounds calculated from the GP approximation and the variational method with a Gaussian wavefunction, respectively. The last one, $E_\mathrm{HP}$, is the Hall-Post lower bound, which is provided by our method as an estimate of the true groundstate energy.}
\label{tab:en}
\end{table}
They were calculated by NLSQ (nonlinear least squares) fitting of the groundstate energy for fixed $N'$ (in our calculations, we have used results for $N'=8$, so as to make $E_\mathrm{HP}$ as high as possible) as function of $L$ to a power law
\[
E(L) = E_\mathrm{HP} + B L^C \ , \quad C < 0
\]
and taking the $L \to \infty$ limit, obtaining $E_\mathrm{HP}$ as the answer. It has been necessary to follow this procedure, since raw numerical results vary with $L$, even for $L$ large enough so that the 1-RDM does not change. The relative asymptotic standard error of the fitting procedure is below 0.01\% for both values of $\lambda$.

The bounds on $E_0$ ($E_\mathrm{HP}$ and $E_\mathrm{GP}$) presented above are quite close. The relative uncertainty with which $E_0$ is determined by them is below 1\% for $\lambda = 10^{-2}$ and around 8\% for $\lambda = 5 \cdot 10^{-2}$. (We have calculated the relative uncertainty as the ratio of the difference between the upper and the lower bound to the lower bound.) The fact that it is small supports the applicability of our approximation, as it means that the true groundstate energy is also close to the obtained lower bound. On the other hand, if the reduced density matrix of the $N'$-particle groundstate $\Psi_0'$  is to be a good approximation of the reduced density matrix of the true groundstate, the mean energy of $\Psi_0'$---i.e. the Hall-Post lower bound---must be close to the true groundstate energy of the system. Our results fulfill this condition.

\subsection{Density matrices}

\begin{figure}[!ht]
\centering\includegraphics[scale=0.7,clip=true,trim=20 0 15 10]{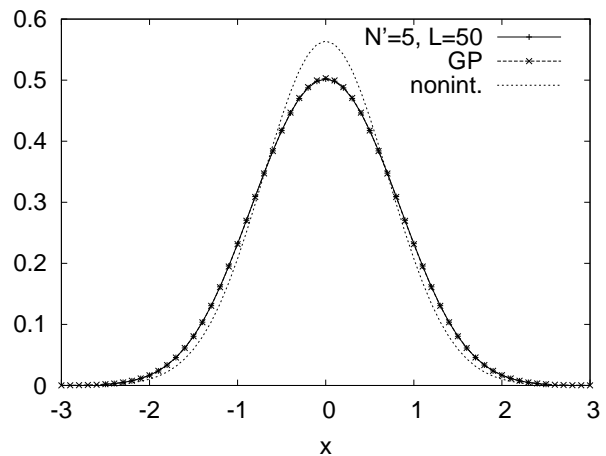}
\caption{Comparison of the probability densities obtained from our method (for $N'=5$ and $L=50$) and from the GP approximation, for $\lambda = 10^{-2}$. The curves overlap perfectly, indicating convergence. The probability density of the noninteracting ($\lambda = 0$) groundstate is shown too. One can clearly see the difference between the interacting system and the noninteracting one, with the repulsive interaction pushing away the bosons in the trap.}
\label{fig:5_50}
\end{figure}

\begin{figure}[!ht]
\centering\includegraphics[scale=0.7,clip=true,trim=20 0 10 5]{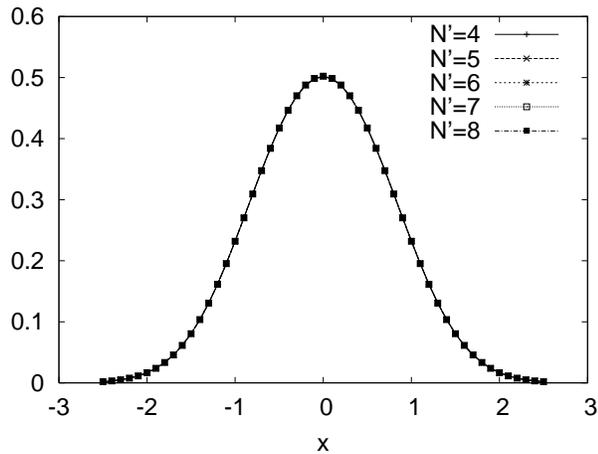}
\caption{Probability densities for $\lambda = 10^{-2}$ calculated for increasing $N'$  ($L=56,50,60,40,40$ respectively). The curves overlap perfectly, indicating convergence.}
\label{fig:conv_1e-2}
\end{figure}

\begin{figure}
\centering\includegraphics[scale=0.7,clip=true,trim=20 0 10 5]{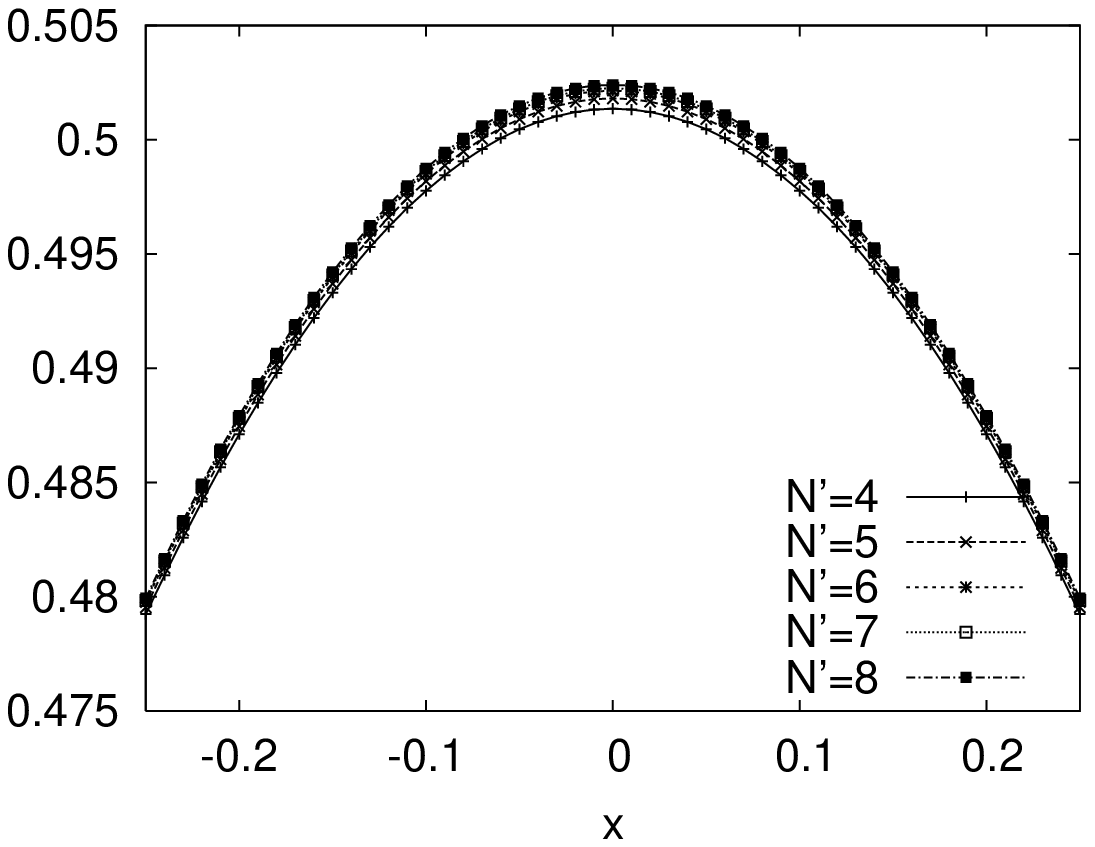}
\caption{Probability densities for $\lambda = 10^{-2}$ calculated for increasing $N'$  ($L=56,50,60,40,40$ respectively), shown in the range $x \in [-0.25,0.25]$. The plot has been maginified in order to show the details of the convergence.}
\label{fig:conv_1e-2m}
\end{figure}

For $\lambda = 10^{-2}$, we obtain identical one-particle probability densities from our method and from the GP approximation, as shown on~Fig.~\ref{fig:5_50}.
The accuracy of our approximation is confirmed by Fig.~\ref{fig:conv_1e-2}, which shows that different values of $N'$ and $L$ yield identical probability densities.
A magnified section of this plot is shown in Fig.~\ref{fig:conv_1e-2m}.
We will use the convergence with increasing $N'$ as a benchmark of the accuracy of our method, treating our numerical results as correct if they stabilize quickly. For each $N'$, we take the results for $L$ large enough so that they do not change upon further increase of $L$. A similar convergence occurs for the antidiagonal part of the 1-RDM, $\rho_1(x,-x)$.

\begin{figure}[!ht]
\centering\includegraphics[scale=0.75,clip=true,trim=40 10 40 10]{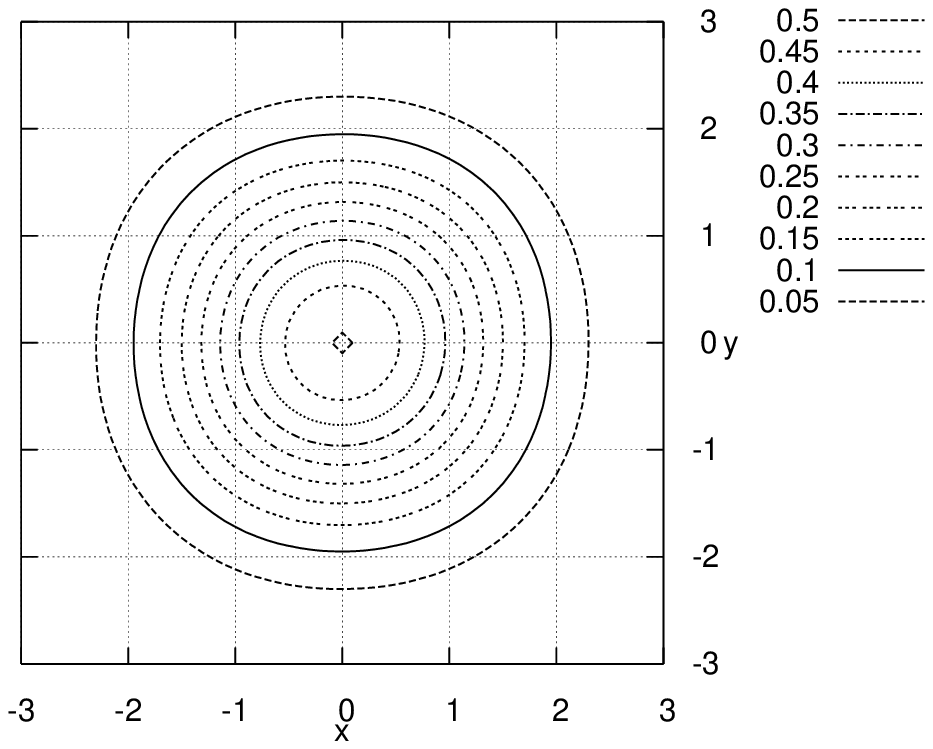}
\caption{Contour plot of $\rho_1(x,y)$ for $\lambda = 10^{-2}$, calculated for $N'=5$ and $L=50$. Isolines display a radial symmetry.}
\label{fig:r1_1e-2_cnt}
\end{figure}

The GP approximation, however, cannot provide us with the knowledge about the nondiagonal parts of the 1-RDM. The merit of our method is that we can calculate $\rho_1(x,y)$ for any $(x,y)$. For $\lambda = 10^{-2}$, we obtain numerically
\begin{equation}
\rho_1(x,y) \approx \rho_1\left(\sqrt{x^2 + y^2} \right) \ ,
\end{equation}
which is clearly shown by the contour plot of $\rho_1(x,y)$ in Fig.~\ref{fig:r1_1e-2_cnt}.

\begin{figure}[!ht]
\centering\includegraphics[scale=0.65]{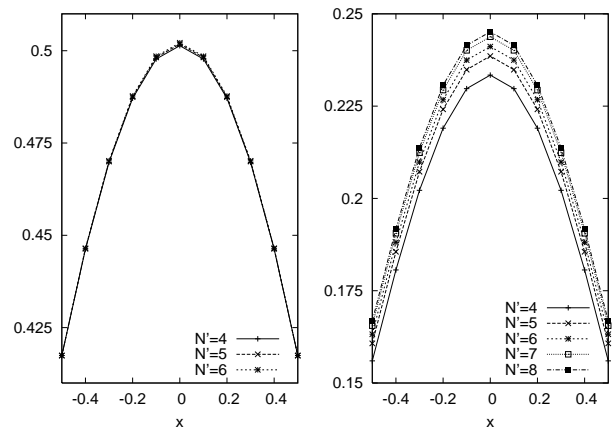}
\caption{On the left plot, the probability density $\rho_1(x,x)$ for $\lambda = 10^{-2}$ is shown for increasing $N'$ ($L=56,50,60$, respectively). The curves converge quickly. On the right plot, the diagonal part of the 2-RDM, $\rho_2(x,x,x,x)$ is plotted (for the same $\lambda$) for increasing $N'$ ($L=56,50,60,40,40$, respectively). Even for $N'$ equal 7 or 8, the curves do not converge.}
\label{fig:1e-2_dd_cmp}
\end{figure}

The convergence of the diagonal part of 1-RDM, $\rho_1(x,x)$ and of the diagonal part of 2-RDM, $\rho_2(x,x,x,x)$ (the probability of finding both particles in the same position $x$), is compared in Fig.~\ref{fig:1e-2_dd_cmp}. It is clear that the convergence, with increasing $N'$, of the second function is much slower. The consequence of this is that using only such simple diagonalization techiques as we did, which limit us to $N' < 10$, we cannot estimate the 2-RDM, even for $\lambda$ as small as $10^{-2}$.

\begin{figure}[!ht]
\centering\includegraphics[scale=0.7,clip=true,trim=10 0 0 0]{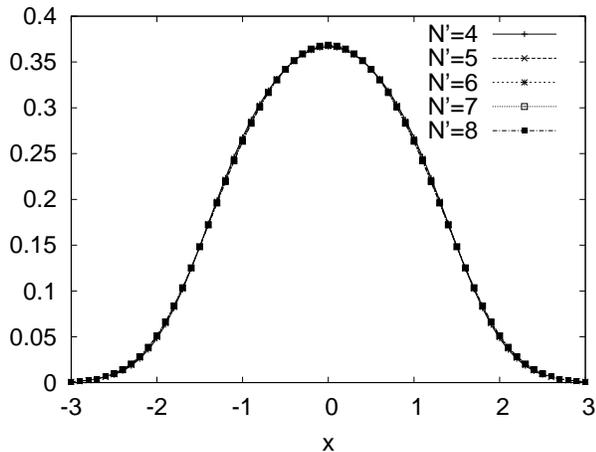}
\caption{Convergence of the probability densities $\rho_1(x,x)$ for $\lambda = 5 \cdot 10^{-2}$ and increasing values of $N'$ ($L = 55, 50, 60, 40, 40$, respectively). The curves overlap, indicating convergence.}
\label{fig:conv_5e-2}
\end{figure}

\begin{figure}
\centering\includegraphics[scale=0.7,clip=true,trim=20 0 10 5]{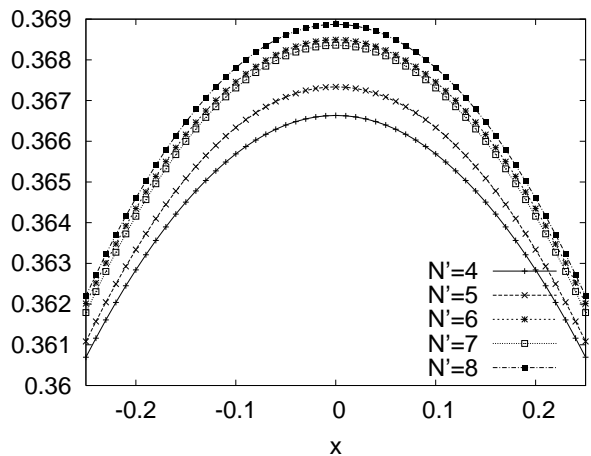}
\caption{Probability densities for $\lambda = 5 \cdot 10^{-2}$ calculated for increasing $N'$  ($L=55,50,60,40,40$ respectively), shown in the range $x \in [-0.25,0.25]$. The plot has been magnified in order to show the details of the convergence.}
\label{fig:conv_5e-2m}
\end{figure}

\begin{figure}[!ht]
\centering\includegraphics[scale=0.7,clip=true,trim=10 0 0 5]{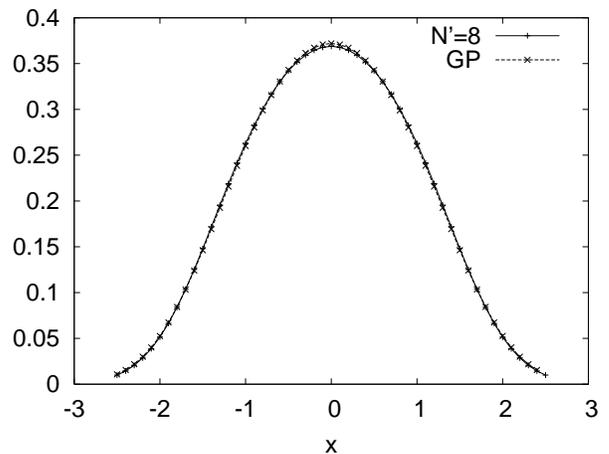}
\caption{Probability density for $\lambda = 5 \cdot 10^{-2}$, as calculated with our method ($N'=8$ and $L=40$) and from the minimization of the GP functional. A slight difference between the two curves can be seen in the middle of the plot, indicating that interactions are strong enough so that GP approximation's results differ from ours.}
\label{fig:prbdens_5e-2}
\end{figure}

\begin{figure}[!ht]
\centering\includegraphics[scale=0.75,clip=true,trim=40 10 40 10]{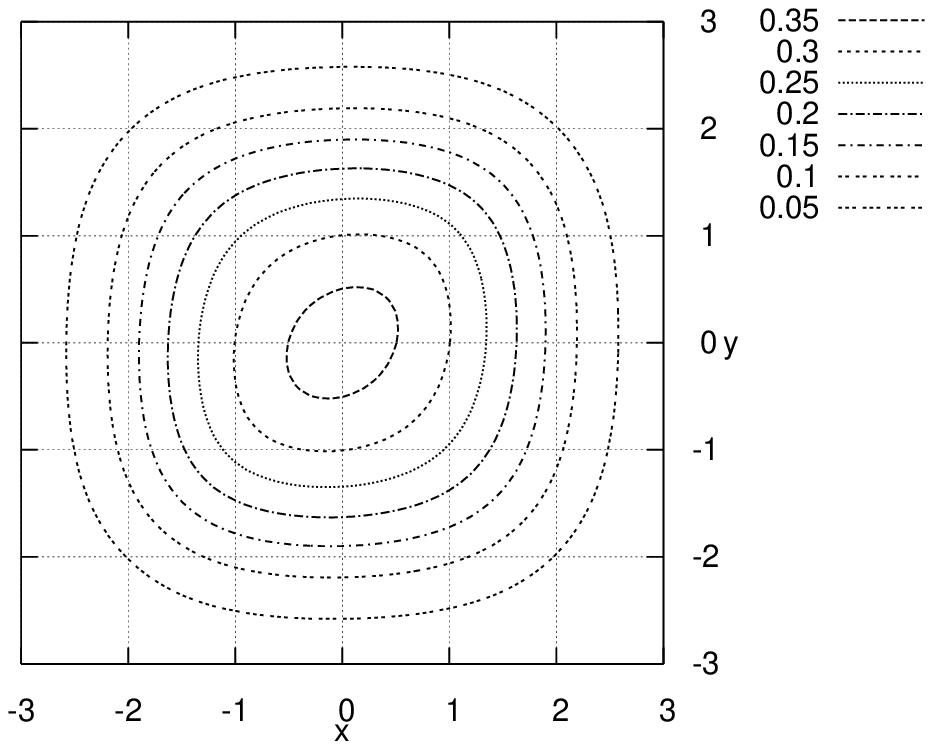}
\caption{Contour plot of $\rho_1(x,y)$ for $\lambda = 5\cdot 10^{-2}$, calculated for $N'=8$ and $L=40$. Isolines do not display the radial symmetry, unlike for $\lambda = 10^{-2}$.}
\label{fig:r1_5e-2_cntr}
\end{figure}

\begin{figure}[!ht]
\centering\includegraphics[scale=0.75,clip=true,trim=40 10 40 10]{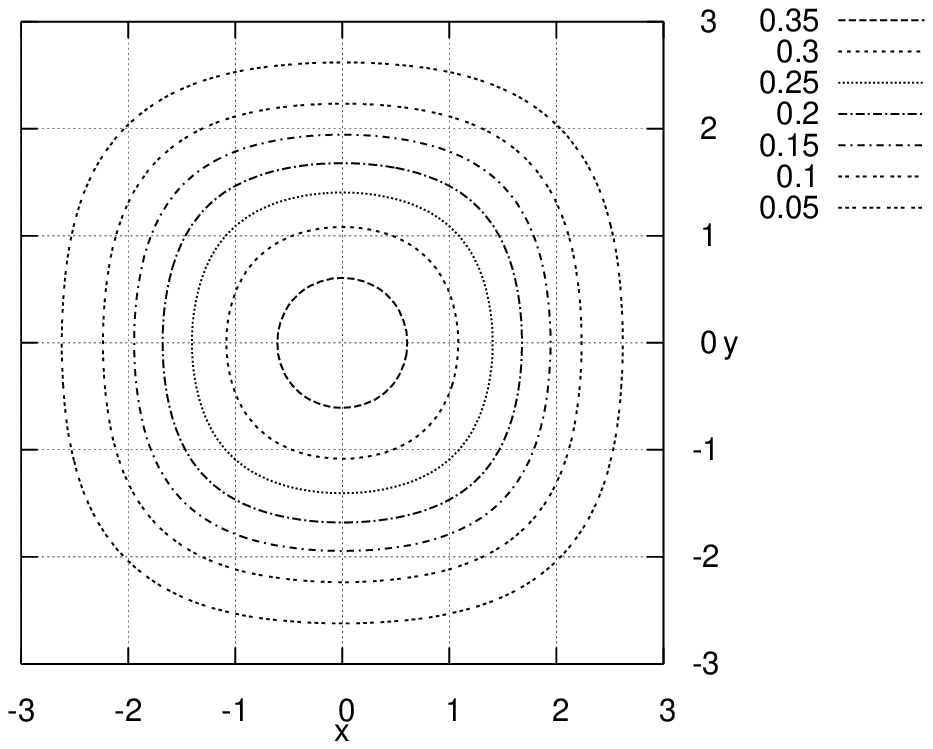}
\caption{Contour plot of $\rho_1(x,y)$ calculated for $\lambda = 5\cdot 10^{-2}$ using GP approximation. It is clearly visible that this plot is more symmetrical than the one in Fig.~\ref{fig:r1_5e-2_cntr}.}
\label{fig:hart}
\end{figure}

For $\lambda = 5 \cdot 10^{-2}$, we obtain the convergence of the probability densities as easily, as for $\lambda = 10^{-2}$ (see Fig.~\ref{fig:conv_5e-2}), although it is slightly slower (not visible on the plot).
A magnified section of this plot is shown in Fig.~\ref{fig:conv_5e-2m}.
A similar convergence occurs for the antidiagonal part of the 1-RDM, $\rho_1(x,-x)$. However, the probability density differs slightly from the one obtained from the GP functional, as seen in Fig.~\ref{fig:prbdens_5e-2}.
Contrary to the case of $\lambda = 10^{-2}$, the contour plot of the 1-RDM for $\lambda = 5 \cdot 10^{-2}$ is no longer radially symmetric, as can be seen on Fig.~\ref{fig:r1_5e-2_cntr}.
It differs noticeably from the one (shown in Fig.~\ref{fig:hart}) we would obtain from the mean field method, using the formula
\[
\rho_1(x,y) \approx \Psi_\mathrm{GP}(x) \Psi_\mathrm{GP}(y) \ ,
\]
where $\Psi_\mathrm{GP}$ is the real wavefunction which minimizes the GP energy functional. This difference is the most striking result in this Section, and indicates that our method gives more accurate results than mean-field approximations.

\begin{figure}[!ht]
\centering\includegraphics[scale=0.65]{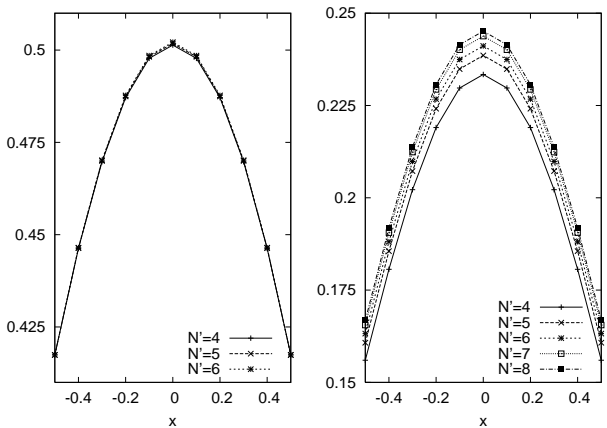}
\caption{On the left plot, the probability density $\rho_1(x,x)$ for $\lambda = 10^{-2}$ is shown for increasing $N'$ ($L=56,50,60$, respectively). On the right plot, the the same quantity is shown for $\lambda = 10^{-1}$ ($L=55,50,60$, respectively). The convergence for the higher value of $\lambda$ is visibly 
slower.}
\label{fig:convcmp}
\end{figure}

\begin{figure}[!ht]
\centering\includegraphics[scale=0.7,clip=true,trim=10 0 0 0]{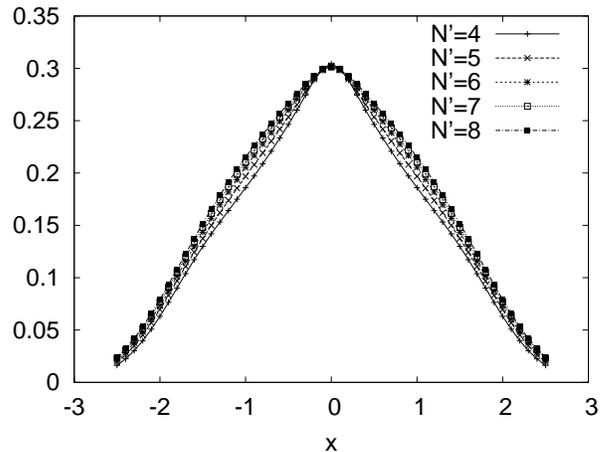}
\caption{Antidiagonal part of 1-RDM for $\lambda = 10^{-1}$ for increasing $N'$ ($L=55,50,60,39,40$, respectively). Convergence is too slow for the results to be meaningful.}
\label{fig:anticonv}
\end{figure}

For even higher value of $\lambda$, $10^{-1}$, we did not obtain fast enough convergence of neither $\rho_1(x,x)$ (see~Fig.~\ref{fig:convcmp})
nor, especially, $\rho_1(x,-x)$ (see Fig.~\ref{fig:anticonv}). This prevented us from investigating the full 1-RDM for this interaction strength.
We conclude that $\lambda = 10^{-1}$ is outside the range of our approximation in its present form. To reach this interaction strength, we would have to calculate the density matrices of the $N'$-particle groundstate for $N'$ higher than those treatable with the simple numerical diagonalization algorithm used by us.

\section{Summary}
\label{sec:summary}

We have presented a method of investigating one- and two-particle reduced density matrices of the groundstate of an interacting systems with a large number of bosonic particles. The method approximates it with a smaller, renormalized system. It is conceptually simple and easy to implement numerically. The results it provides would be, for high enough interaction strengths (e.g. $\lambda = 5 \cdot 10^{-2}$ in our example), impossible to calculate using mean-field methods, such as the GP approximation.

We have provided an example application of our method to the problem of one-dimensional interacting Bose gas in a harmonic trap, obtaining accurate approximations of the quantity unobtainable from mean field methods, namely the full one-particle density matrix. The results are precise and accurately describe the large system, which is proven by the fact the the results converge quickly with increasing $N'$. The GP approximation does not give as accurate picture of the groundstate one-particle density matrix as our approach. Additionally, the Hall-Post lower bounds for the groundstate energy have been calculated. Relatively small difference between them and the upper bounds (GP and Gaussian) also supports the applicability of our method.

Even using simple numerical procedures, our method gives access to properties which were previously not as accurately described by mean-field methods. To investigate the two-particle density matrix, or to perform simulations of systems with higher interaction strengths, would require the use of more refined approaches to the calculation of the groundstate of the renormalized Hamiltonian, for example the DMRG method~\cite{white1,schollwock} or the variational optimization of the basis wavefunctions~\cite{spalek1,spalek2,spalek3}.

\begin{acknowledgments}
The authors want to thank prof.~Iwo~Bia\l ynicki-Birula and prof.~Kazimierz~Rz\c{a}\.{z}ewski for helpful discussions and advice.
\end{acknowledgments}

%\bibliography{blockgrnd}

\begin{thebibliography}{18}
\expandafter\ifx\csname natexlab\endcsname\relax\def\natexlab#1{#1}\fi
\expandafter\ifx\csname bibnamefont\endcsname\relax
  \def\bibnamefont#1{#1}\fi
\expandafter\ifx\csname bibfnamefont\endcsname\relax
  \def\bibfnamefont#1{#1}\fi
\expandafter\ifx\csname citenamefont\endcsname\relax
  \def\citenamefont#1{#1}\fi
\expandafter\ifx\csname url\endcsname\relax
  \def\url#1{\texttt{#1}}\fi
\expandafter\ifx\csname urlprefix\endcsname\relax\def\urlprefix{URL }\fi
\providecommand{\bibinfo}[2]{#2}
\providecommand{\eprint}[2][]{\url{#2}}

\bibitem[{\citenamefont{Pitaevskii and Stringari}(2003)}]{pitstr}
\bibinfo{author}{\bibfnamefont{L.~P.} \bibnamefont{Pitaevskii}}
  \bibnamefont{and}
  \bibinfo{author}{\bibfnamefont{S.}~\bibnamefont{Stringari}},
  \emph{\bibinfo{title}{Bose-Einstein Condensation}}
  (\bibinfo{publisher}{Oxford University Press}, \bibinfo{address}{Oxford},
  \bibinfo{year}{2003}).

\bibitem[{\citenamefont{Dalfovo et~al.}(1999)\citenamefont{Dalfovo, Giorgini,
  Pitaevskii, and Stringari}}]{pitaevskii2}
\bibinfo{author}{\bibfnamefont{F.}~\bibnamefont{Dalfovo}},
  \bibinfo{author}{\bibfnamefont{S.}~\bibnamefont{Giorgini}},
  \bibinfo{author}{\bibfnamefont{L.~P.} \bibnamefont{Pitaevskii}},
  \bibnamefont{and}
  \bibinfo{author}{\bibfnamefont{S.}~\bibnamefont{Stringari}},
  \bibinfo{journal}{Rev. Mod. Phys.} \textbf{\bibinfo{volume}{71}},
  \bibinfo{pages}{463} (\bibinfo{year}{1999}).

\bibitem[{\citenamefont{White}(1992)}]{white1}
\bibinfo{author}{\bibfnamefont{S.~R.} \bibnamefont{White}},
  \bibinfo{journal}{Phys. Rev. Lett.} \textbf{\bibinfo{volume}{69}},
  \bibinfo{pages}{2863} (\bibinfo{year}{1992}).

\bibitem[{\citenamefont{Schollw\"{o}ck}(2005)}]{schollwock}
\bibinfo{author}{\bibfnamefont{U.}~\bibnamefont{Schollw\"{o}ck}},
  \bibinfo{journal}{Rev. Mod. Phys.} \textbf{\bibinfo{volume}{77}},
  \bibinfo{pages}{259} (\bibinfo{year}{2005}).

\bibitem[{\citenamefont{Tonks}(1936)}]{tonks}
\bibinfo{author}{\bibfnamefont{L.}~\bibnamefont{Tonks}},
  \bibinfo{journal}{Phys. Rev.} \textbf{\bibinfo{volume}{50}},
  \bibinfo{pages}{955} (\bibinfo{year}{1936}).

\bibitem[{\citenamefont{Girardeau}(1960)}]{girardeau}
\bibinfo{author}{\bibfnamefont{M.}~\bibnamefont{Girardeau}},
  \bibinfo{journal}{J. Math. Phys.} \textbf{\bibinfo{volume}{1}},
  \bibinfo{pages}{516} (\bibinfo{year}{1960}).

\bibitem[{\citenamefont{Spa{\l}ek et~al.}(2000)\citenamefont{Spa{\l}ek,
  Podsiad{\l}y, W\'{o}jcik, and Rycerz}}]{spalek1}
\bibinfo{author}{\bibfnamefont{J.}~\bibnamefont{Spa{\l}ek}},
  \bibinfo{author}{\bibfnamefont{R.}~\bibnamefont{Podsiad{\l}y}},
  \bibinfo{author}{\bibfnamefont{W.}~\bibnamefont{W\'{o}jcik}},
  \bibnamefont{and} \bibinfo{author}{\bibfnamefont{A.}~\bibnamefont{Rycerz}},
  \bibinfo{journal}{Phys. Rev. B} \textbf{\bibinfo{volume}{61}},
  \bibinfo{pages}{15676} (\bibinfo{year}{2000}).

\bibitem[{\citenamefont{Spa{\l}ek and Rycerz}(2001{\natexlab{a}})}]{spalek2}
\bibinfo{author}{\bibfnamefont{J.}~\bibnamefont{Spa{\l}ek}} \bibnamefont{and}
  \bibinfo{author}{\bibfnamefont{A.}~\bibnamefont{Rycerz}},
  \bibinfo{journal}{Phys. Rev. B} \textbf{\bibinfo{volume}{63}},
  \bibinfo{pages}{073101} (\bibinfo{year}{2001}{\natexlab{a}}).

\bibitem[{\citenamefont{Spa{\l}ek and Rycerz}(2001{\natexlab{b}})}]{spalek3}
\bibinfo{author}{\bibfnamefont{J.}~\bibnamefont{Spa{\l}ek}} \bibnamefont{and}
  \bibinfo{author}{\bibfnamefont{A.}~\bibnamefont{Rycerz}},
  \bibinfo{journal}{Phys. Rev. B} \textbf{\bibinfo{volume}{64}},
  \bibinfo{pages}{161105(R)} (\bibinfo{year}{2001}{\natexlab{b}}).

\bibitem[{\citenamefont{Lieb and Liniger}(1963)}]{lieb}
\bibinfo{author}{\bibfnamefont{E.~H.} \bibnamefont{Lieb}} \bibnamefont{and}
  \bibinfo{author}{\bibfnamefont{W.}~\bibnamefont{Liniger}},
  \bibinfo{journal}{Phys. Rev.} \textbf{\bibinfo{volume}{130}},
  \bibinfo{pages}{1605} (\bibinfo{year}{1963}).

\bibitem[{\citenamefont{Kheruntsyan et~al.}(2003)\citenamefont{Kheruntsyan,
  Gangardt, Drummond, and Shlyapnikov}}]{shlyapnikov2}
\bibinfo{author}{\bibfnamefont{K.~V.} \bibnamefont{Kheruntsyan}},
  \bibinfo{author}{\bibfnamefont{D.~M.} \bibnamefont{Gangardt}},
  \bibinfo{author}{\bibfnamefont{P.~D.} \bibnamefont{Drummond}},
  \bibnamefont{and} \bibinfo{author}{\bibfnamefont{G.~V.}
  \bibnamefont{Shlyapnikov}}, \bibinfo{journal}{Phys. Rev. Lett.}
  \textbf{\bibinfo{volume}{91}}, \bibinfo{pages}{040403}
  (\bibinfo{year}{2003}).

\bibitem[{\citenamefont{Kheruntsyan et~al.}(2005)\citenamefont{Kheruntsyan,
  Gangardt, Drummond, and Shlyapnikov}}]{shlyapnikov3}
\bibinfo{author}{\bibfnamefont{K.~V.} \bibnamefont{Kheruntsyan}},
  \bibinfo{author}{\bibfnamefont{D.~M.} \bibnamefont{Gangardt}},
  \bibinfo{author}{\bibfnamefont{P.~D.} \bibnamefont{Drummond}},
  \bibnamefont{and} \bibinfo{author}{\bibfnamefont{G.~V.}
  \bibnamefont{Shlyapnikov}}, \bibinfo{journal}{Phys. Rev. A}
  \textbf{\bibinfo{volume}{71}}, \bibinfo{pages}{053615}
  (\bibinfo{year}{2005}).

\bibitem[{\citenamefont{Gangardt and Shlyapnikov}(2003)}]{shlyapnikov}
\bibinfo{author}{\bibfnamefont{D.~M.} \bibnamefont{Gangardt}} \bibnamefont{and}
  \bibinfo{author}{\bibfnamefont{G.~V.} \bibnamefont{Shlyapnikov}},
  \bibinfo{journal}{Phys. Rev. Lett.} \textbf{\bibinfo{volume}{90}},
  \bibinfo{pages}{010401} (\bibinfo{year}{2003}).

\bibitem[{\citenamefont{Coleman}(1963)}]{coleman}
\bibinfo{author}{\bibfnamefont{A.~J.} \bibnamefont{Coleman}},
  \bibinfo{journal}{Rev. Mod. Phys.} \textbf{\bibinfo{volume}{35}},
  \bibinfo{pages}{668} (\bibinfo{year}{1963}).

\bibitem[{\citenamefont{Ruskai}(1969)}]{ruskai}
\bibinfo{author}{\bibfnamefont{M.~B.} \bibnamefont{Ruskai}},
  \bibinfo{journal}{Phys. Rev.} \textbf{\bibinfo{volume}{183}},
  \bibinfo{pages}{129} (\bibinfo{year}{1969}).

\bibitem[{\citenamefont{Gidofalvi and Mazziotti}(2004)}]{gidofalvi}
\bibinfo{author}{\bibfnamefont{G.}~\bibnamefont{Gidofalvi}} \bibnamefont{and}
  \bibinfo{author}{\bibfnamefont{D.~A.} \bibnamefont{Mazziotti}},
  \bibinfo{journal}{Phys. Rev. A} \textbf{\bibinfo{volume}{69}},
  \bibinfo{pages}{042511} (\bibinfo{year}{2004}).

\bibitem[{\citenamefont{Hall and Post}(1967)}]{hallpost}
\bibinfo{author}{\bibfnamefont{R.~L.} \bibnamefont{Hall}} \bibnamefont{and}
  \bibinfo{author}{\bibfnamefont{H.~R.} \bibnamefont{Post}},
  \bibinfo{journal}{Proc. Phys. Soc.} \textbf{\bibinfo{volume}{90}},
  \bibinfo{pages}{381} (\bibinfo{year}{1967}).

\bibitem[{\citenamefont{Lehoucq et~al.}()\citenamefont{Lehoucq, Maschhoff,
  Sorensen, and Yang}}]{arpack}
\bibinfo{author}{\bibfnamefont{R.}~\bibnamefont{Lehoucq}},
  \bibinfo{author}{\bibfnamefont{K.}~\bibnamefont{Maschhoff}},
  \bibinfo{author}{\bibfnamefont{D.}~\bibnamefont{Sorensen}}, \bibnamefont{and}
  \bibinfo{author}{\bibfnamefont{C.}~\bibnamefont{Yang}},
  \emph{\bibinfo{title}{\textsc{ARPACK} -- {A}rnoldi package}},
  \bibinfo{howpublished}{http://www.caam.rice.edu/software/ARPACK/}.

\end{thebibliography}
%\bibliographystyle{apsrev}

\end{document}